\documentclass{article}
\usepackage{spconf,amsmath,graphicx,color,glossaries,siunitx,pgfplots,
tabularx,booktabs,multirow,url,balance}
\usepackage{tikz}
\graphicspath{{images/}}

\newacronym{AI}{AI}{artificial intelligence}
\newacronym{AP}{AP}{average precision}
\newacronym{ASC}{ASC}{acoustic scene classification}
\newacronym{AUC}{AUC}{area under curve}
\newacronym{CRNN}{CRNN}{convolutional recurrent neural network}
\newacronym{CT}{CT}{cross trigger}
\newacronym{CTTC}{CTTC}{cross trigger tolerance criterion}
\newacronym{DCASE}{DCASE}{Detection and Classification of Acoustic Scenes and Events}
\newacronym{DTC}{DTC}{detection tolerance criterion}
\newacronym{eFPR}{eFPR}{effective FPR}
\newacronym{eTPR}{eTPR}{effective TPR}
\newacronym{ER}{ER}{error rate}
\newacronym{FBCRNN}{FBCRNN}{forward-backward convolutional recurrent neural network}
\newacronym{FN}{FN}{false negative}
\newacronym{FP}{FP}{false positive}
\newacronym{FPR}{FPR}{FP-Rate}
\newacronym{GTC}{GTC}{ground truth intersection criterion}
\newacronym{PR}{PR}{precision-recall}
\newacronym{PSD}{PSD}{polyphonic sound detection}
\newacronym{PSDS}{PSDS}{PSD score}
\newacronym{ROC}{ROC}{receiver operating characteristic}
\newacronym{SED}{SED}{sound event detection}
\newacronym{TCCRNN}{TCCRNN}{tag-conditioned CRNN}
\newacronym{TP}{TP}{true positive}
\newacronym{TPR}{TPR}{TP-Rate}
\title{Threshold Independent Evaluation of Sound Event Detection Scores}
\twoauthors
  {Janek Ebbers, Reinhold Haeb-Umbach\thanks{Funded by the Deutsche Forschungsgemeinschaft (DFG, German Research Foundation) - 282835863.}}
	{Paderborn University,\\
	Department of Communications Engineering,\\
	33098 Paderborn, Germany,\\\{ebbers,haeb\}@nt.upb.de}
  {Romain Serizel}
	{Universit\'{e} de Lorraine, CNRS,\\
	Inria, Loria,\\
	F-54000 Nancy, France,\\
	romain.serizel@loria.fr}

\begin{document}
\ninept
\maketitle
\begin{abstract}
\vspace{-1mm}
Performing an adequate evaluation of \gls{SED} systems is far from trivial and is still subject to ongoing research.
The recently proposed \gls{PSD}-\gls{ROC} and \gls{PSDS} make an important step into the direction of an evaluation of \gls{SED} systems which is independent from a certain decision threshold.
This allows to obtain a more complete picture of the overall system behavior which is less biased by threshold tuning.
Yet, the \gls{PSD}-\gls{ROC} is currently only approximated using a finite set of thresholds.
The choice of the thresholds used in approximation, however, can have a severe impact on the resulting \gls{PSDS}.
In this paper we propose a method which allows for computing system performance on an evaluation set for all possible thresholds jointly, enabling accurate computation not only of the \gls{PSD}-\gls{ROC} and \gls{PSDS} but also of other collar-based and intersection-based performance curves.
It further allows to select the threshold which best fulfills the requirements of a given application.
Source code is publicly available in our \gls{SED} evaluation package \textit{sed\_scores\_eval}\footnote{\url{https://github.com/fgnt/sed_scores_eval}}.
\end{abstract}
\vspace{-1mm}

\vspace{-1mm}
\section{Introduction}
\label{sec:intro}
\vspace{-1mm}
\glsresetall
Recently, there is a rapid progress in Machine Listening aiming to imitate by machines the human ability to recognize, distinguish and interpret sounds~\cite{virtanen2018computational}.
The progress is driven by the annual \gls{DCASE} challenges\footnote{\url{http://dcase.community/events#challenges}} and the releases of large-scale sound databases such as Google's AudioSet~\cite{gemmeke2017audio} and FSD50k~\cite{fonseca2020fsd50k}.

For a successful development of such systems an adequate evaluation of the system's operating behavior is crucial, where, ideally, the evaluation metric correlates to the user satisfaction during system application~\cite{krstulovic2018audio}.

In this paper we are concerned with the evaluation of \gls{SED} systems~\cite{mesaros2021sound}.
\Gls{SED} aims to recognize sound events in audio signals together with their onset and offset time.
One particular challenge in \gls{SED} is that labeling of ground truth event onset and offset times, referred to as strong labels, is expensive and time-consuming.
Therefore, many systems aim to learn \gls{SED} from weakly labeled data~\cite{shah2018closer,miyazaki2020weakly}, which only indicate the presence or absence of a sound event in an audio signal without providing its onset and offset times, and unlabeled data~\cite{Lu2018,turpault2020training}.
Syntheticly generated soundscapes are another alternative to produce cheap strongly annotated data~\cite{turpault2021sound,ronchini2021impact}.
Here, an insightful evaluation of systems is particularly important to be able to draw conclusions about the system's learning behavior w.r.t. the temporal localization of sounds.

Due to the temporal component of sound events, however, the adequate evaluation of \gls{SED} performance is far from trivial.
Traditional approaches perform segment-based and collar-based (event-based) evaluation~\cite{mesaros2016metrics} for only a single operating point (decision threshold).
Further, segment-based evaluation does not sufficiently evaluate a system's capability of providing connected detections, whereas collar-based evaluation is sensitive to ambiguities in the definition of the ground truth event boundaries.

More recently, Bilen et al. \cite{bilen2020framework} proposed the \gls{PSD}-\gls{ROC} curve and \gls{PSDS}, which is an important step towards an evaluation of \gls{SED} systems which is independent of specific decision thresholds and therefore provides a more complete picture of the system's overall operating behavior and is less biased by a specific tuning of the decision thresholds.

However, \gls{PSD}-\gls{ROC} curves are only approximated so far due to the lack of a method which efficiently evaluates the system's performance for all possible decision thresholds.
The approximation of the \gls{PSD}-\gls{ROC} curve can significantly underestimate the system's \gls{PSDS} as we will show in Sec.~\ref{sec:experiments}.

In this paper, we therefore present such a method to efficiently compute the system's performance for all possible decision thresholds jointly, which allows us to accurately compute the \gls{PSD}-\gls{ROC} and \gls{PSDS}.
Further, it can also be used to compute other intersection-based and collar-based performance curves such as \gls{PR}-curves.
The presented approach can be understood as a generalization of the method used for single instance evaluation\footnote{By single instance evaluation we refer to an evaluation where each classified instance is evaluated with its own target.} to more sophisticated evaluations such as collar-based or intersection-based evaluations.
It is based on the definition of changes in the intermediate statistics that occur when the decision threshold falls below a certain score, which we refer to as deltas in the following.
Then, absolute values can be obtained for all possible thresholds by performing a cumulative sum over the deltas.

The rest of the paper is structured as follows.
Sec.~\ref{sec:sed_eval} reviews current threshold-dependent approaches for \gls{SED} evaluation.
Sec.~\ref{sec:thres_ind_eval} describes commonly used threshold-independent evaluation methods for single instance evaluation\footnotemark[3] as well as the recently proposed \gls{PSD} for the threshold-independent evaluation of \gls{SED}.
Then, we present our proposed approach for the accurate computation of \gls{PSD}-\gls{ROC} and other performance curves in Sec.~\ref{sec:implementation}.
Finally we present experiments in Sec.~\ref{sec:experiments} and draw conclusions in Sec.~\ref{sec:conclusions}.

%
%The \gls{DCASE} Challenge Task~4 combines all three types of data for several years now~\cite{Serizel2018,turpault2019sound,turpault2020training,turpault2021sound,ronchini2021impact} with the aim of exploiting all kinds of available data sources.

\section{Sound Event Detection Evaluation}
\label{sec:sed_eval}
\Gls{SED}~\cite{virtanen2018computational,mesaros2021sound} can be seen as a multi-label classification problem, where the system performs classifications at multiple points in time which usually happens in a frame-based manner.
When a classification score $y_t$ exceeds a certain decision threshold it is marked as positive.
Connected positive classifications are merged into a detected event $(\hat{t}_{\text{on},i}, \hat{t}_{\text{off},i}, \hat{c}_i)$ with $\hat{t}_{\text{on},i}, \hat{t}_{\text{off},i}, \hat{c}_i$ being the onset time, offset time and class label, respectively, of the $i$-th detection.

As in other classification tasks the evaluation is based on \gls{TP}, \gls{FP} and \gls{FN} counts.
The \glspl{TP} count $N_{\text{TP}}$ represents the number of ground truth events that have been detected by the system.
The \glspl{FP} count $N_{\text{FP}}$ sums up the number of detections which do not match a ground truth event.
Hence, the total number of detected events is given as $N_\text{DP}=N_\text{TP}+N_\text{FP}$.
The \glspl{FN} count $N_\text{FN}$, which is the number of ground truth events missed by the system, is given as $N_\text{FN}=N_\text{GP}-N_\text{TP}$ with $N_\text{GP}$ being the total number of ground truth events.
From these intermediate statistics higher level measures can be derived such as the precision $P=N_\text{TP}/N_\text{DP}$, the recall (\glspl{TPR}) $R=N_\text{TP}/N_\text{GP}$ and \gls{FPR} $\text{FPR}=N_\text{FP}/N_\text{GN}$, where $N_\text{GN}$ is the total number of ground truth negative instances in the evaluation data set.

Compared to single instance evaluation\footnotemark[3], it is less obvious in \gls{SED} when to classify a ground truth event as detected, i.e. \gls{TP}, and when to consider a detection as \gls{FP}, due to the temporal extent of the target events over multiple classification scores/frames.
Currently there exist three conceptually different ways for this, which are segment-based, collar-based (event-based) and intersection-based \cite{mesaros2016metrics,mesaros2018datasets,bilen2020framework,ferroni2021improving}.

\subsection{Segment-based}
\vspace{-1mm}
In segment-based evaluation~\cite{mesaros2016metrics,mesaros2018datasets}, classifications and targets are defined in fixed length segments (\SI{1}{s} segments is a popular choice).
Classifications and targets are considered positive if they are detected/labeled anywhere in the segment.
This way evaluation can be treated as a single instance evaluation.
However, segment-based evaluation overemphasizes the contribution of longer events which expand over multiple segments and it does not evaluate the system's capability of providing meaningful uninterrupted detections.
%Therefore, in 2018 the DCASE Challenge Task~4 moved from segment-based to collar-based evaluation, which is explained next.

\subsection{Collar-based}
\vspace{-1mm}
Collar-based, a.k.a. event-based, evaluation~\cite{mesaros2016metrics,mesaros2018datasets} compares detections $(\hat{t}_{\text{on},i}, \hat{t}_{\text{off},i}, \hat{c}_i)$ with ground truth events $(t_{\text{on},j}, t_{\text{off},j}, c_j)$ directly.
Only if there is a matching event pair $(i,j)$ with $c_j=\hat{c}_i$, $|\hat{t}_{\text{on},i}-t_{\text{on},j}|\leq d$ and $|\hat{t}_{\text{off},i}-t_{\text{off},j}|\leq d_{\text{off},j}$, a \gls{TP} is achieved.
Other detections are counted as \glspl{FP}.
The offset collar $d_{\text{off},j}=\mathrm{max}(d,rT_j)$ usually depends on the length $T_j$ of the ground truth event.
Common choices are $d=\SI{200}{ms}$ and $r=0.2$.

With collar-based evaluation, each ground truth event has equal contribution to the overall performance and systems can only achieve good performance if events are detected as single connected detections.
This, however, introduces sensitivity to ambiguities in the annotation.
If, e.g., an annotator labeled multiple dog barks as a single event but a system detects each bark as a separate event, this results in multiple \glspl{FP} and one \gls{FN}.
%Therefore, the intersection-based evaluation has been proposed recently \cite{} as explained in the following Section.

\subsection{Intersection-based}
\vspace{-1mm}
Intersection-based evaluation~\cite{bilen2020framework,ferroni2021improving} determines the number of \glspl{TP} and \glspl{FP} based on intersections between detections and ground truth events.
A \gls{DTC} classifies detections as \gls{FP} if its intersection with ground truth events of the same event class, normalized by the length of the detected event, falls below a certain \gls{DTC} ratio $\rho_\text{DTC}$.
Else, it is considered relevant, which, however, does not necessarily mean \gls{TP}.
A ground truth event is only classified \gls{TP} if its intersection with relevant same class detections, normalized by the length of the ground truth event, is greater or equal to a \gls{GTC}~ratio~$\rho_\text{GTC}$.
%Note that if, e.g., there is only a single detection lying completely within a ground truth event but only covers \SI{10}{\%} of it, it is neither a \gls{TP} nor a \gls{FP}.

Bilen et al. \cite{bilen2020framework} further introduced \glspl{CT} which are \gls{FP} detections matching events from another event class and, thus, may impair user experience more than standalone \glspl{FP}.
Note that, although the concept of \glspl{CT} has been proposed in conjunction with intersection-based evaluation, it is not restricted to it and could also be transferred to segment-based and collar-based evaluations.
In intersection-based evaluation the \gls{CTTC} counts a \gls{CT} between a detected event class $\hat{c}_i$ and another event class $c$ with $c\neq \hat{c}_i$ if the detection intersects with ground truth events of class $c$ by at least $\rho_\text{CTTC}$.

\vspace{-2mm}
\section{Threshold-Independent Evaluation}
\label{sec:thres_ind_eval}
\vspace{-2mm}
The computation of above intermediate statistics, such as the \gls{TP} count, depend on the decision threshold that is applied to the classifier's output scores.
Consequently, metrics such as $F_1$-scores and error-rates only evaluate a single threshold.
A more complete picture of the classifier's performance, however, can be obtained when evaluating system performance for all possible thresholds.

\subsection{Single Instance Evaluation}
\label{sec:single_inst_curves}
\vspace{-1mm}
\begin{figure}[t]
  \centering
  \def\svgwidth{0.9\linewidth}
  {\footnotesize
  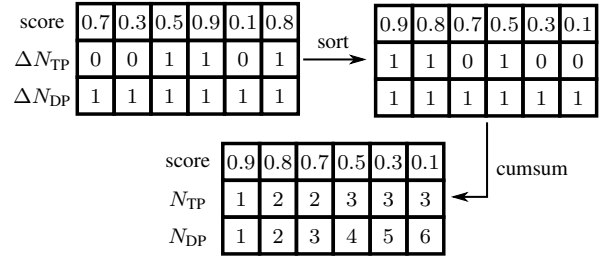
  }
  \vspace{-1mm}
  \caption{Illustration of the joint computation of intermediate statistics with single instance evaluation.}
  \vspace{-3mm}
  \label{fig:score_sort}
\end{figure}

In single instance evaluation\footnotemark[3], the \gls{PR} and \gls{ROC} curves~\cite{Davis06therelationship,mesaros2018datasets} are frequently used to evaluate overall system behavior independently from a certain operating point.
As the name suggests, the \gls{PR} curve plots precisions over corresponding recall values which result from arbitrary decision thresholds.
The \gls{ROC} curve instead plots the recalls over corresponding \glspl{FPR}.
Frequently used metrics for system comparison are the area under the \gls{PR} curve, a.k.a. \gls{AP}, and the area under the \gls{ROC} curve, which is often simply referred to as \gls{AUC}.

Rather than making decisions and evaluating performance seperately for a set of arbitrary thresholds, performance can be evaluated for all thresholds jointly by implementing a sorting of classification scores $y$ together with some predefined deltas, as it is done, e.g., in the \textit{scikit-learn toolkit}~\cite{scikit-learn}. 
Here, deltas mean changes in the intermediate statistics, such as the number of \glspl{TP}, when the decision threshold moves from above an instance's classification score to below of it, i.e., when the instance moves from being classified negative to being classified positive.
Then absolute values can be obtained by simply performing a cumulative sum of the deltas.

This approach is illustrated in Fig.~\ref{fig:score_sort} for an exemplary data set with six instances.
$\Delta N_\text{TP}$ means the change in the \gls{TP} count which, for single instance evaluation, is simply the binary target of the instance.
This is because, upon positive classification, the \gls{TP} count only increases by one when the instance is labeled positive.
$\Delta N_\text{DP}$ represents the change in the total number of system detections.
Here $\Delta N_\text{DP}$ is always one as there is always one instance more being classified positive when the threshold falls below its classification score.
The precisions $P=N_\text{TP}/N_\text{DP}$ can, e.g., now be read off for all decision thresholds in the third table containing the absolute values. 

\subsection{PSD-ROC}
\vspace{-1mm}
To the best of our knowledge, the \gls{PSD}-\gls{ROC} curve proposed in~\cite{bilen2020framework} is currently the only threshold-independent evaluation of \gls{SED} systems. %\inred{It has been used for the evaluation of the DCASE 2021 submissions ...}
It first computes, for all event classes $c$, intersection-based \gls{ROC} curves $\text{ROC}_c(\text{eFPR})$ which are monotonically increasing curves plotting \gls{TPR} over \gls{eFPR}, where the reader is referred to Bilen et al.~\cite{bilen2020framework} for further details about its computation.
%,where the \gls{eFPR} for an event class $c$ is defined as
%\begin{align}
%\mathrm{eFPR}_c = \frac{N_{\text{FP},c}}{T_\text{ds}} + \alpha_\text{CT}\frac{1}{C-1}\sum_{\substack{m\\m\neq c}}\frac{N_{\text{CT},c,m}}{\sum_{j\in\mathcal{G}_m}T_j}\,\,\, .
%\end{align}
%Here, $N_{\text{FP},c}$ is the number of \glspl{FP} of event class $c$, $T_\text{ds}$ is the total duration of the evaluation dataset, $\alpha_\text{CT}$ is the cost of \glspl{CT} on user experience, $C$ is the number of event classes, $N_{\text{CT},c,m}$ is the number of cross triggers between predicted events of class $c$ and ground truth events of class~$m$ and $\mathcal{G}_m$ is the set of ground truth event indices from events of class~$m$, respectively.
The final \gls{PSD}-\gls{ROC} summarizes the classwise \gls{ROC} curves as
\begin{align}
\text{PSD-ROC}(\text{eFPR}) = \mu_\text{TPR}(\text{eFPR}) - \alpha_\text{ST}\cdot \sigma_\text{TPR}(\text{eFPR})\, ,
\end{align}
with $\mu_\text{TPR}(\text{eFPR})$ and $\sigma_\text{TPR}(\text{eFPR})$ being the mean and standard deviation over the classwise ROC curves at a certain \gls{eFPR}, and where $\alpha_\text{ST}$ is a parameter penalizing instability across classes.
The \gls{PSDS} is the normalized area under the PSD-ROC curve up to a maximal $\text{eFPR}_\text{max}$.

Note that the number of thresholds, which may result in a different \gls{TPR}-\gls{eFPR} value pair, is as high as the number of classification scores in the data set.
With a system outputting scores at a rate of \SI{50}{Hz} and a rather small evaluation set of, e.g., only \SI{1}{h}, this would be $\SI{180}{k}$ thresholds to be evaluated for each event class.
Evaluating system performance for each of the thresholds separately is not feasible for obvious reasons.
Therefore, due to a lack of an efficient joint computation of intersection-based \gls{TPR}-\gls{eFPR} value pairs for all thresholds, the PSD-ROC curve is commonly approximated with a reduced set of thresholds.
For instance, the DCASE 2021 Challenge Task~4~\cite{ronchini2021impact} employed \glspl{PSDS} using 50 linearly spaced thresholds.
The approximation of PSD-ROC curves, however, can lead to a significant underestimation of the \gls{PSDS} as we will demonstrate in Sec.~\ref{sec:experiments}.
Non-linearly spaced thresholds could alleviate this to some extent, which, however, remains arbitrary and ad-hoc.

\vspace{-2mm}
\section{Efficient Computation of Collar- and Intersection-based Curves}
\label{sec:implementation}
\vspace{-2mm}
\begin{figure}[t]
  \centering
  \def\svgwidth{0.71\linewidth}
  {\footnotesize
  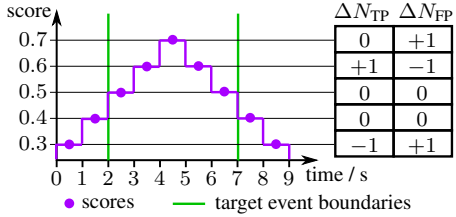
  }
  \vspace{-2mm}
  \caption{Collar-based deltas example.}
  \vspace{-3mm}
  \label{fig:deltas_cb}
\end{figure}

\begin{figure}[t]
  \centering
  \def\svgwidth{0.8\linewidth}
  {\footnotesize
  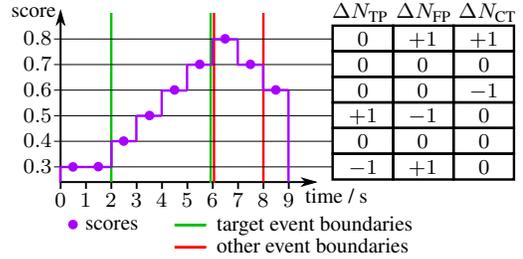
  }
  \vspace{-2mm}
  \caption{Intersection-based deltas example.}
  \vspace{-3mm}
  \label{fig:deltas_ib}
\end{figure}

\begin{figure*}[t]
    \newlength\figureheight
    \newlength\figurewidth
    \setlength\figureheight{3.1cm}
    \setlength\figurewidth{6cm}
  % This file was created by tikzplotlib v0.9.8.
\begin{tikzpicture}

\definecolor{color0}{rgb}{0.12156862745098,0.466666666666667,0.705882352941177}

\begin{axis}[
title={50 thresholds},
height=\figureheight,
width=\figurewidth,
tick align=outside,
tick pos=left,
x grid style={white!69.0196078431373!black},
xlabel={eFPR per hour},
xmajorgrids,
xmin=0, xmax=100,
xtick style={color=black},
y grid style={white!69.0196078431373!black},
ylabel={eTPR},
ymajorgrids,
ymin=0, ymax=0.5,
ytick style={color=black}
]
\addplot [very thick, red, const plot mark right]
table {%
0 0
2.09994477922988 0
2.62493097403735 0
3.14991716884482 0
3.67490336365229 0
4.19988955845976 0
4.72487575326723 0
5.2498619480747 0
5.77484814288217 0
6.29983433768964 0.0471287360963337
6.82482053249711 0.139328216834322
7.34980672730458 0.145300565871318
7.87479292211205 0.145380789934903
8.39977911691952 0.15188106228422
8.92476531172699 0.15188106228422
9.44975150653446 0.15188106228422
9.97473770134193 0.15188106228422
10.4997238961494 0.151992798226397
11.0247100909569 0.152084145120234
11.5496962857643 0.152084145120234
12.0746824805718 0.264498406940407
14.6996134546092 0.264498406940407
15.2245996494166 0.264498406940407
16.2745720390316 0.264498406940407
16.799558233839 0.264498406940407
17.3245444286465 0.264498406940407
17.849530623454 0.265974286623464
18.3745168182614 0.265974286623464
18.8995030130689 0.265974286623464
19.4244892078764 0.265974286623464
19.9494754026839 0.265974286623464
20.4744615974913 0.265974286623464
23.6243787663361 0.265974286623464
24.1493649611436 0.26624623125396
24.6743511559511 0.379966087758628
26.774295935181 0.379966087758628
27.2992821299884 0.380063454967483
27.8242683247959 0.380063454967483
28.3492545196034 0.380063454967483
28.8742407144108 0.380117110172807
29.3992269092183 0.380117110172807
30.4491992988333 0.380117110172807
31.4991716884482 0.380117110172807
32.0241578832557 0.380117110172807
32.5491440780631 0.380117110172807
33.0741302728706 0.380117110172807
34.1241026624855 0.380117110172807
34.649088857293 0.380167205817485
35.1740750521005 0.380167205817485
36.2240474417154 0.380167205817485
36.7490336365229 0.421027570647357
38.3239922209453 0.421027570647357
38.8489784157528 0.421027570647357
39.3739646105602 0.421027570647357
41.4739093897901 0.421027570647357
41.9988955845976 0.421027570647357
42.5238817794051 0.421027570647357
44.6238265586349 0.421027570647357
45.1488127534424 0.421027570647357
45.6737989482499 0.421027570647357
46.1987851430573 0.421027570647357
46.7237713378648 0.421027570647357
47.2487575326723 0.421027570647357
47.7737437274798 0.421027570647357
48.8237161170947 0.421027570647357
49.3487023119022 0.421027570647357
50.9236608963246 0.421027570647357
51.9736332859395 0.421027570647357
53.5485918703619 0.421027570647357
54.5985642599769 0.421027570647357
55.1235504547843 0.421027570647357
55.6485366495918 0.421027570647357
56.1735228443993 0.421027570647357
57.7484814288217 0.429232915469238
58.7984538184366 0.429232915469238
59.8484262080516 0.429232915469238
60.373412402859 0.429232915469238
61.423384792474 0.429232915469238
61.9483709872815 0.429232915469238
64.0483157665113 0.429232915469238
64.5733019613188 0.429232915469238
65.0982881561263 0.431268742698754
66.1482605457412 0.431268742698754
66.6732467405487 0.431268742698754
67.1982329353561 0.431268742698754
67.7232191301636 0.431268742698754
68.7731915197786 0.431268742698754
69.298177714586 0.431268742698754
70.348150104201 0.431268742698754
70.8731362990084 0.431268742698754
71.3981224938159 0.431268742698754
72.4480948834308 0.431268742698754
73.4980672730458 0.431268742698754
74.0230534678533 0.431268742698754
74.5480396626607 0.431268742698754
75.0730258574682 0.431268742698754
75.5980120522757 0.431268742698754
77.1729706366981 0.431268742698754
78.222943026313 0.431268742698754
78.7479292211205 0.431268742698754
79.272915415928 0.431268742698754
81.8978463899653 0.431268742698754
82.4228325847728 0.431268742698754
82.9478187795802 0.431268742698754
83.4728049743877 0.431268742698754
83.9977911691952 0.431268742698754
84.5227773640026 0.431268742698754
86.0977359484251 0.431268742698754
86.6227221432325 0.431268742698754
87.14770833804 0.431268742698754
87.6726945328475 0.431268742698754
88.7226669224624 0.431268742698754
89.2476531172699 0.431268742698754
90.2976255068848 0.431268742698754
90.8226117016923 0.431268742698754
91.8725840913072 0.431268742698754
92.3975702861147 0.431268742698754
92.9225564809222 0.431268742698754
93.4475426757296 0.431268742698754
95.0225012601521 0.431268742698754
95.5474874549595 0.431268742698754
96.5974598445745 0.431268742698754
97.1224460393819 0.431268742698754
97.6474322341894 0.431268742698754
98.1724184289969 0.431268742698754
98.6974046238043 0.431268742698754
99.2223908186118 0.431268742698754
99.7473770134193 0.431268742698754
100.272363208227 0.431268742698754
101.322335597842 0.431268742698754
101.847321792649 0.431268742698754
102.372307987457 0.431268742698754
103.947266571879 0.431268742698754
104.472252766687 0.431268742698754
104.997238961494 0.431268742698754
105.522225156301 0.431268742698754
106.047211351109 0.431268742698754
106.572197545916 0.431268742698754
107.097183740724 0.431268742698754
107.622169935531 0.431268742698754
108.147156130339 0.431268742698754
108.672142325146 0.431268742698754
109.197128519954 0.431268742698754
109.722114714761 0.431268742698754
110.247100909569 0.431268742698754
110.772087104376 0.431268742698754
122.846769584948 0.431268742698754
146.471148351284 0.431268742698754
182.170209598192 0.431268742698754
};
\addplot [very thick, color0, const plot mark right]
table {%
0 -0.0695846747934266
0.52498619480747 -0.0288161833387389
1.04997238961494 -0.00538910385113064
1.57495858442241 0.0430050090902352
2.09994477922988 0.0929930345534758
2.62493097403735 0.0998092109696785
3.14991716884482 0.134404190632213
3.67490336365229 0.150877554089084
4.19988955845976 0.159355572448253
4.72487575326723 0.164585051109893
5.2498619480747 0.200071652481656
5.77484814288217 0.244169117835469
6.29983433768964 0.310717785119752
6.82482053249711 0.333940596804827
7.34980672730458 0.342853999181215
7.87479292211205 0.352550531085183
8.39977911691952 0.354125225844478
8.92476531172699 0.356894846192866
9.44975150653446 0.361317837395815
9.97473770134193 0.368167237829534
10.4997238961494 0.372060247758882
11.0247100909569 0.375430078963183
11.5496962857643 0.377183360496638
12.0746824805718 0.378694895586221
12.5996686753793 0.379851096716234
13.1246548701867 0.382547549881156
14.1746272598017 0.383509064411719
14.6996134546092 0.385282589323752
15.2245996494166 0.385547357697514
15.7495858442241 0.385690005264941
16.2745720390316 0.38773636056126
16.799558233839 0.388724473959645
17.849530623454 0.388987540031934
18.3745168182614 0.396568507759317
18.8995030130689 0.398369533640788
19.4244892078764 0.400827630626424
19.9494754026839 0.40236296408486
20.9994477922988 0.402751993520605
21.5244339871063 0.407131188001794
22.0494201819137 0.408202741635286
22.5744063767212 0.408896407723532
23.0993925715287 0.408910202092933
23.6243787663361 0.412228118950358
24.6743511559511 0.413214479741935
25.724323545566 0.413570133889572
26.2493097403735 0.413610829230059
27.2992821299884 0.413623525257337
29.9242131040258 0.413647617008531
30.4491992988333 0.413851477268986
35.1740750521005 0.416633054705306
36.2240474417154 0.424261687973659
37.2740198313304 0.429053327028178
38.8489784157528 0.429735127225147
43.0488679742125 0.430416234032252
48.2987299222872 0.431096642875943
49.3487023119022 0.431776349149154
65.0982881561263 0.433133635386953
100 0.433133635386953
};
\end{axis}

\end{tikzpicture}
  \input{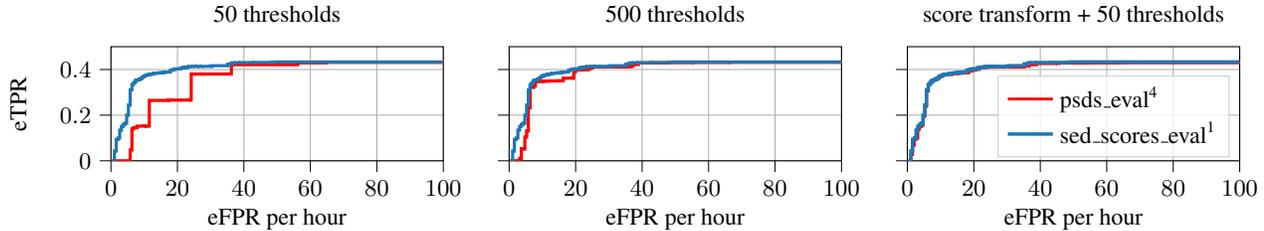}
  % This file was created by tikzplotlib v0.9.8.
\begin{tikzpicture}

\definecolor{color0}{rgb}{0.12156862745098,0.466666666666667,0.705882352941177}

\begin{axis}[
title={score transform + 50 thresholds},
title style={align=center},
height=\figureheight,
width=\figurewidth,
legend cell align={left},
legend style={
  fill opacity=0.8,
  draw opacity=1,
  text opacity=1,
  at={(0.97,0.03)},
  anchor=south east,
  draw=white!80!black
},
tick align=outside,
tick pos=left,
x grid style={white!69.0196078431373!black},
xlabel={eFPR per hour},
xmajorgrids,
xmin=0, xmax=100,
xtick style={color=black},
y grid style={white!69.0196078431373!black},
ymajorgrids,
ymin=0, ymax=0.5,
yticklabels=\empty,
ytick style={color=black}
]
\addplot [line width=1.08pt, red, const plot mark right]
table {%
0 0
0.52498619480747 0
1.04997238961494 0
1.57495858442241 0.026982347819919
2.09994477922988 0.0682886674950593
2.62493097403735 0.09289665563937
3.14991716884482 0.0984636893629818
3.67490336365229 0.138580428263161
4.19988955845976 0.149408379230865
4.72487575326723 0.155086351924543
5.2498619480747 0.194170412885351
5.77484814288217 0.238303648164482
6.29983433768964 0.305541830819522
6.82482053249711 0.334903313945939
7.34980672730458 0.343724983045275
7.87479292211205 0.34847326677933
8.39977911691952 0.354770554697854
8.92476531172699 0.359993293259434
9.44975150653446 0.361686271516577
9.97473770134193 0.369995782011378
10.4997238961494 0.369995782011378
11.0247100909569 0.377589243320012
11.5496962857643 0.377835278125754
12.0746824805718 0.378144751647429
13.1246548701867 0.378144751647429
13.6496410649942 0.378503502513201
14.6996134546092 0.382074817277141
15.2245996494166 0.382305796178397
15.7495858442241 0.382632203623669
16.2745720390316 0.382632203623669
17.3245444286465 0.384661320098611
18.3745168182614 0.384661320098611
18.8995030130689 0.391505521851039
19.4244892078764 0.394820639502136
19.9494754026839 0.396008825076315
20.4744615974913 0.397180244966979
21.5244339871063 0.399847766744
22.0494201819137 0.405397288580531
23.6243787663361 0.408424452504339
24.1493649611436 0.408424452504339
25.724323545566 0.410216349847128
26.2493097403735 0.410216349847128
26.774295935181 0.410306966276575
28.3492545196034 0.410306966276575
28.8742407144108 0.410328954212899
30.9741854936407 0.410328954212899
31.4991716884482 0.410328954212899
32.0241578832557 0.410328954212899
32.5491440780631 0.410328954212899
33.0741302728706 0.410328954212899
33.5991164676781 0.410328954212899
34.1241026624855 0.410367755940279
35.1740750521005 0.410367755940279
36.7490336365229 0.413839766169922
37.2740198313304 0.420049961908928
37.7990060261378 0.420049961908928
38.8489784157528 0.420049961908928
39.3739646105602 0.425525496047741
43.57385416902 0.425525496047741
44.6238265586349 0.425525496047741
45.6737989482499 0.426206865754291
47.2487575326723 0.427567507283865
47.7737437274798 0.427567507283865
49.3487023119022 0.427567507283865
53.0236056755545 0.427567507283865
54.5985642599769 0.427567507283865
55.1235504547843 0.427567507283865
55.6485366495918 0.428246769843919
58.2734676236292 0.428246769843919
59.8484262080516 0.428246769843919
60.373412402859 0.428246769843919
62.4733571820889 0.428246769843919
64.0483157665113 0.428246769843919
65.0982881561263 0.428925320702155
66.1482605457412 0.428925320702155
67.1982329353561 0.428925320702155
68.7731915197786 0.428925320702155
69.298177714586 0.428925320702155
69.8231639093935 0.428925320702155
70.8731362990084 0.428925320702155
74.5480396626607 0.428925320702155
75.0730258574682 0.428925320702155
76.1229982470831 0.428925320702155
77.6979568315056 0.428925320702155
81.3728601951578 0.428925320702155
82.4228325847728 0.428925320702155
82.9478187795802 0.428925320702155
83.4728049743877 0.428925320702155
83.9977911691952 0.428925320702155
84.5227773640026 0.428925320702155
87.6726945328475 0.428925320702155
88.1976807276549 0.428925320702155
89.7726393120774 0.428925320702155
91.8725840913072 0.428925320702155
92.9225564809222 0.428925320702155
95.5474874549595 0.428925320702155
96.5974598445745 0.428925320702155
97.1224460393819 0.428925320702155
98.1724184289969 0.428925320702155
99.2223908186118 0.428925320702155
100.272363208227 0.428925320702155
101.847321792649 0.428925320702155
102.372307987457 0.428925320702155
104.997238961494 0.428925320702155
105.522225156301 0.428925320702155
106.572197545916 0.428925320702155
108.147156130339 0.428925320702155
109.197128519954 0.428925320702155
110.247100909569 0.428925320702155
118.121893831681 0.428925320702155
156.445886052626 0.428925320702155
207.894533143758 0.428925320702155
};
\addlegendentry{psds\_eval\footnotemark[4]}
\addplot [line width=1.32pt, color0, const plot mark right]
table {%
0 -0.0695846747934266
0.52498619480747 -0.0288161833387389
1.04997238961494 -0.00538910385113064
1.57495858442241 0.0430050090902352
2.09994477922988 0.0929930345534758
2.62493097403735 0.0998092109696785
3.14991716884482 0.134404190632213
3.67490336365229 0.150877554089084
4.19988955845976 0.159355572448253
4.72487575326723 0.164585051109893
5.2498619480747 0.200071652481656
5.77484814288217 0.244169117835469
6.29983433768964 0.310717785119752
6.82482053249711 0.333940596804827
7.34980672730458 0.342853999181215
7.87479292211205 0.352550531085183
8.39977911691952 0.354125225844478
8.92476531172699 0.356894846192866
9.44975150653446 0.361317837395815
9.97473770134193 0.368167237829534
10.4997238961494 0.372060247758882
11.0247100909569 0.375430078963183
11.5496962857643 0.377183360496638
12.0746824805718 0.378694895586221
12.5996686753793 0.379851096716234
13.1246548701867 0.382547549881156
14.1746272598017 0.383509064411719
14.6996134546092 0.385282589323752
15.2245996494166 0.385547357697514
15.7495858442241 0.385690005264941
16.2745720390316 0.38773636056126
16.799558233839 0.388724473959645
17.849530623454 0.388987540031934
18.3745168182614 0.396568507759317
18.8995030130689 0.398369533640788
19.4244892078764 0.400827630626424
19.9494754026839 0.40236296408486
20.9994477922988 0.402751993520605
21.5244339871063 0.407131188001794
22.0494201819137 0.408202741635286
22.5744063767212 0.408896407723532
23.0993925715287 0.408910202092933
23.6243787663361 0.412228118950358
24.6743511559511 0.413214479741935
25.724323545566 0.413570133889572
26.2493097403735 0.413610829230059
27.2992821299884 0.413623525257337
29.9242131040258 0.413647617008531
30.4491992988333 0.413851477268986
35.1740750521005 0.416633054705306
36.2240474417154 0.424261687973659
37.2740198313304 0.429053327028178
38.8489784157528 0.429735127225147
43.0488679742125 0.430416234032252
48.2987299222872 0.431096642875943
49.3487023119022 0.431776349149154
65.0982881561263 0.433133635386953
100 0.433133635386953
};
\addlegendentry{sed\_scores\_eval\footnotemark[1]}
\end{axis}

\end{tikzpicture}
  \vspace{-3mm}
  \caption{PSD-ROC curves: The exact PSD-ROC curve being shown in blue, which becomes computable with our proposed methodology, and different approximations of the PSD-ROC curve shown in red.}
  \vspace{-3mm}
  \label{fig:psd_rocs}
\end{figure*}
In this section we present how collar-based and intersection-based intermediate statistics can be efficiently computed jointly for all possible decision thresholds.
For this we follow the same approach used for the computation of single instance evaluation curves which we described in Sec.~\ref{sec:single_inst_curves}.
We aim to bring all classification scores into a sorted list together with the deltas of the intermediate statistics, which appear when the decision threshold falls below the classification score.
Then we are able to obtain absolute values for all operating points by a simple cumulative sum over the deltas.

With collar-based and intersection-based evaluation, however, the computation of the deltas becomes more challenging compared to single instance evaluation, as here all scores of an audio signal have to be considered jointly and cannot be obtained instance-wise.
The basic principle of the definition of the deltas is illustrated in Fig.~\ref{fig:deltas_cb} and Fig.~\ref{fig:deltas_ib}.

In Fig.~\ref{fig:deltas_cb} collar-based evaluation is considered.
For simplicity, we here assume scores/frames to have a width of \SI{1}{s}, that target event boundaries lie exactly between two scores/frames and the {on-/off}set collars to be \SI{1}{s}.
Starting from a decision threshold above $0.7$, no event would be detected as no score lies above the threshold.
When the decision threshold falls below $0.7$, a detection is spawned from second 4 to 5 as the 5th score lies above the threshold.
However, the distances between the detected and the true onsets and offsets are \SI{2}{s} for both, therefore not matching the collar.
Hence, the newly spawned detection is a \gls{FP} and we have $\Delta N_\text{FP} = +1$.
When the threshold falls below $0.6$, however, the detection expands from second $3$ to $6$ and the \gls{FP} disappears ($\Delta N_\text{FP}= -1$) and becomes a \gls{TP} detection ($\Delta N_\text{TP}= +1$).
When the decision threshold falls below $0.5$ and below $0.4$, nothing changes as the collars are still matched and the detection remains \gls{TP} ($\Delta N_\text{TP}=\Delta N_\text{FP}=0$).
Finally, when the decision threshold falls below $0.3$, the detection expands from \SI{0}{s} to \SI{9}{s} and the detected on-/offsets exceed the collar, and the \gls{TP} disappears ($\Delta N_\text{TP}=-1$) and becomes a \gls{FP} again~($\Delta N_\text{FP}=+1$).

A slightly more advanced example is shown in Fig.~\ref{fig:deltas_ib}, where we consider intersection-based evaluation including \glspl{CT}.
We assume $\rho_\text{DTC}=\rho_\text{GTC}=\rho_\text{CTTC}=0.5$ and that again all event boundaries lie exactly between two scores/frames.
When the decision threshold falls below $0.8$ here, a detection is spawned from \SI{6}{s} to \SI{7}{s} which does not overlap with the target event at all, giving us $\Delta N_\text{FP}=+1$.
Further, the detected event completely lies within the ground truth event from another class (in red), giving us $\Delta N_\text{CT}=+1$.
When the threshold falls below $0.7$, the detection's overlap with the target event is still only $1/3<\rho_\text{DTC}$.
This is still a \gls{FP} and therefore $\Delta N_\text{FP}=0$.
The overlap with the other class event is $2/3\geq\rho_\text{CTTC}$. Therefore there is still a \gls{CT}, with $\Delta N_\text{CT}=0$.
When the threshold falls below $0.6$, the detection's overlap with both the target event and the other class event is $2/5<\rho_\text{DTC}=\rho_\text{CTTC}$.
The detection is still \gls{FP} ($\Delta N_\text{FP}=0$), but not a \gls{CT} anymore ($\Delta N_\text{CT}=-1$).
When the threshold falls below $0.5$ the overlap with the target event becomes $1/2=\rho_\text{DTC}$.
The \gls{FP} disappears ($\Delta N_\text{FP}=-1$) and becomes a \gls{TP} ($\Delta N_\text{TP}=+1$).
This remains unchanged until the decision threshold falls below $0.3$, where the overlap with the ground truth event becomes only $4/9<\rho_\text{DTC}$.
This is a \gls{FP} again (but not a \gls{CT}) with $\Delta N_\text{TP}=-1$ and $\Delta N_\text{FP}=+1$.

The proposed approach allows for efficient and accurate computation of collar-based and intersection-based \gls{PR} and \gls{ROC} curves, which not only enables us to compute threshold-independent metrics such as \gls{AP} and \gls{PSDS} precisely, but it also allows us to find the threshold which best suits specific application requirements.

Note that the proposed methodology is rather general and can be applied to arbitrary evaluations as long as one is able to determine the deltas in the intermediate statistics for each classification score in the evaluation data set.

\vspace{-2mm}
\section{Experiments}
\label{sec:experiments}
\vspace{-2mm}
In this section we demonstrate the usefulness of the proposed method for the accurate computation of threshold-independent curves and metrics as well as its potential for threshold tuning.

The presented curves and metrics are evaluated for one of our single model systems developed for DCASE 2021 Challenge Task~4, which employs a \gls{FBCRNN} for audio tagging followed by a \gls{TCCRNN} for \gls{SED}~\cite{Ebbers2021} outputting detection scores at a rate of \SI{50}{Hz}.
For more details about the system and its training, which are not relevant here, the reader is referred to Ebbers et al.~\cite{Ebbers2021}.

In the challenge, systems have been evaluated by \glspl{PSDS} which have been calculated using 50 thresholds linearly spaced from $0.01$ to $0.99$ for \gls{PSD}-\gls{ROC} curve approximation.
In the following we consider the scenario 1 with $\rho_{DTC}=\rho_{GTC}=0.7$, $\alpha_\text{CT}=0$, $\alpha_\text{ST}=1$ and $\text{eFPR}_\text{max}=100/\text{h}$ and report evaluations on the public evaluation set of the DESED database~\cite{turpault2019sound}.

In Fig.~\ref{fig:psd_rocs} different \gls{PSD}-\gls{ROC} curves are shown.
In the subplots we present different variants of \gls{PSD}-\gls{ROC} curve approximations (in red), which have been generated using the official \textit{psds\_eval} package\footnote{\url{https://github.com/audioanalytic/psds_eval}}, and compare them with the accurate \gls{PSD}-\gls{ROC} curve (in blue), which has been generated with our newly released package \textit{sed\_scores\_eval}\footnotemark[1].

During our system development for the challenge, we recognized that our system mostly produces either very small or very high scores, which, without further measures, results in the \gls{PSD}-\gls{ROC} being approximated only very coarsely as shown in the left subplot of Fig.~\ref{fig:psd_rocs}.
Compared to the accurate computation proposed here, the approximated \gls{PSDS} of $0.358$ significantly underestimates the true \gls{PSDS} of $0.400$.
Even if 500 linearly spaced thresholds from $0.001$ to $0.999$ are used, which is shown in the middle plot, this ``step'' artifact still appears on the PSD-ROC. The PSDS computed with these thresholds results to be $0.389$ which still underestimates the true \gls{PSDS}.

In order to obtain a smooth \gls{PSD}-\gls{ROC} in the challenge, we performed a non-linear transformation of our system's classification scores, such that the classification scores of ground truth positive frames in the validation set are uniformly distributed between $0$ and $1$.
Note, that a non-linear score transformation followed by linearly spaced thresholds results to be the same as non-linearly spaced thresholds.
The resulting PSD-ROC approximation with $50$ thresholds is shown in red in the right plot of Fig.~\ref{fig:psd_rocs}, which then comes close to the true PSD-ROC.
Note, that at this point a tuning of a score transformation function (or alternatively 50 thresholds) is required, which is highly undesired for a supposedly threshold-independent metric.
However, with the proposed computation approach, the \gls{PSDS} can be computed exactly and truly independently of a specific set of thresholds (with less computation time\footnote{See \url{https://github.com/fgnt/sed_scores_eval/blob/main/notebooks/psds.ipynb} for timings.}).
%Furthermore, it is computationally more efficient than evaluating performance for many thresholds, separately.

Next, we use the collar-based PR-curve to perform optimal threshold tuning for collar-based $F_1$-score evaluation, which  has been an additional contrastive metric in the challenge.
For each event class we choose the decision threshold, which achieves the highest $F_1$-score on the PR-curve of the validation set that was computed with the proposed approach.
Table~\ref{tab:fscore-results} shows collar-based $F_1$-score performance on the public evaluation set comparing the threshold, which is optimal on the validation set, with simply choosing a threshold of $0.5$.
Note that for a fair comparison, we performed a median filter size sweep for each threshold variant separately and chose for each threshold variant and event class the filter size that performed best on the validation set.
At this point it may be worth noting that median filtering before and after a thresholding yields the same detection outputs, making it similarly applicable to \gls{SED} scores before computing threshold-independent curves or metrics.

\begin{table}[t]
	\vspace{-2mm}
	\caption{Collar-based $F_1$-score performance without and with optimal threshold tuning on validation set.}
	\vspace{0.5mm}
	\label{tab:fscore-results}
	\centering
	\begin{tabular}{c|cc}
		\multirow{2}{*}{Thresholds} & \multirow{2}{*}{$0.5$} & optimal \\
		 & & (on val. set) \\
  		\noalign{\hrule height 1pt}
		$F_1$-score & $\SI{51.8}{\%}$ & $\SI{57.2}{\%}$ \\
	\end{tabular}
	\vspace{-4mm}
\end{table}

It can be observed that solely by tuning the decision threshold on the validation set, performance can be improved by $\SI{5.4}{\%}$.
This demonstrates how threshold-dependent metrics can be biased by the tuning of an operating point.
However, it also demonstrates the ability of our presented method to allow for searching the optimal operating point for a given target application.

\vspace{-2mm}
\section{Conclusions}
\label{sec:conclusions}
\vspace{-2mm}
In this paper we presented a methodology allowing for performing accurate computation of collar-based and intersection-based \gls{PR} and \gls{ROC} curves.
Computing these metrics on a fixed set of thresholds could lead to biased estimation of the final metric.
This can result in significant performance underestimation if an unfavorable set of thresholds is chosen.
Our  proposed method, however, enables truly threshold-independent collar-based and intersection-based SED metrics and provides a more accurate, system independent evaluation.
Further, as the method allows to efficiently compute performances for arbitrary thresholds, it allows to determine the best operating point to fulfill the requirements of a specific application.
We publicly released its implementation in a python package termed \textit{sed\_scores\_eval}\footnotemark[1].

\balance

\bibliographystyle{IEEEbib}
\bibliography{refs}

\end{document}